\documentclass[prb, onecolumn]{revtex4-2} 

\usepackage{paracol}

\usepackage{amsmath}  
\usepackage{amsfonts} 
\usepackage{graphicx} 
\usepackage{xcolor}
\def\diag{\mathop{\mathrm{diag}}}

\begin{document}


\title{General relativity and background Lorentz transformations solve Supplee's submarine paradox}

\author{Osama Karkout} 
\affiliation{} 

\begin{abstract}
A submarine moving at relativistic horizontal velocity sinks in Earth's rest frame due to length contraction while appearing to float in its own frame. Using spacetime geometry and the Lorentz transformations, we show that the resolution lies in how metric components transform between reference frames in relative motion. This solution frees us from assumptions made in previous studies on how a Newtonian gravitational force should transform. The method of background Lorentz transformations is technically simpler than previous treatments in the framework of general relativity. Moreover, we find a novel and intuitive understanding of the paradox, and correct an erroneous expression for the gravitational force obtained by Supplee and used again in the literature.
\end{abstract}

\maketitle 
\section{Introduction and motivation}

It is common in physics to introduce theoretical paradoxes and try to resolve them to gain deeper insight into the theory.
First proposed by James M. Supplee in 1989, Supplee's submarine paradox challenges our understanding of relativity.
Consider a submarine of rest density $\rho_0$ moving in a horizontal direction in an idealized ocean: the water is static, has density $\rho_0$, and has no hydrodynamic effects such as drag, turbulence, or viscosity.
In the rest frame of the ocean, the submarine undergoes Lorentz length contraction and has density $\gamma\rho_0$, higher than that of the water. According to the classical buoyancy principle, the submarine will sink.
At first glance, there seems to be a paradox when we move to the submarine’s rest frame: the submarine’s density here is $\rho_0$. The water volume element moves and undergoes Lorentz length contraction, so its density is $\gamma\rho_0$. This is higher than the submarine's, resulting in the submarine floating instead of sinking.
Since the submarine cannot float in one frame and sink in another, we reach an apparent contradiction.

Supplee \cite{Sup} first avoided the complications that gravity brings to the problem by evoking the equivalence principle. An exercise in special relativity shows that the ocean's floor is no longer flat in a freely falling frame inside the submarine, and the submarine sinks relative to the ocean floor in both reference frames. Supplee then uses a formula for the gravitational force acting upon a moving object in the rest frame of the earth \cite{Landau}. Incidentally, that formula is valid only if the object moves at a small velocity \cite{Landau}, so using it in this context is wrong. In what follows we derive the appropriate expression for the force. Regardless, Supplee then shows that the resulting downward acceleration matches his previous calculations for the ocean's rest frame, which by itself is not a second solution to the paradox. Rather, it is a hint that Supplee's method within special relativity gives inaccurate expressions, and we can correct them for high velocities using general relativity.

With the idea that a general-relativistic analysis is required for a more rigorous solution, a subsequent study by George E. A. Matsas \cite{Mat} sets up the problem in a Rindler spacetime, starts with a submarine initially at rest, and studies its rigidity as it accelerates to a fixed horizontal velocity. By modeling the water as a perfect fluid, Matsas calculates the proper hydrostatic pressure on the top and bottom of the submarine after the acceleration phase, then reaches an expression for a net force downwards acting on the submarine, making it sink. Matsas shows that we can minimize shear effects and turbulence by making the submarine thin enough compared to the vertical pressure gradient, which gives us confidence in neglecting hydrodynamic effects.

Although rigorous and elaborate, this study sidesteps the essence of the paradox; in principle, one should be able to find an answer to Supplee's original question without considering an acceleration phase, since a contradiction arises even for a submarine initially moving at a constant velocity. With this in mind, we set out to resolve the paradox using general relativity but with the initial conditions of a submarine in uniform horizontal motion. We're lucky that the solution we find is simpler and brings additional insight.

Another study by R. S. Vieira \cite{Vie} found the same inappropriate expression used by Supplee for the gravitational force because it starts from the same metric \cite{Landau} used to derive the formula. Vieira then returns to special relativity, assumes this force to be Lorentz covariant, and finds its expression in the submarine's frame. As Matsas noted, within the context of special relativity, assumptions about how the Newtonian gravitational field would transform in different reference frames are unavoidable. Moreover, it is natural to seek a solution in the more general theory. As it stands, the paradox presents an opportunity to probe our understanding of general relativity, and that is what we set out to do.
In section \ref{coo} we briefly present needed formulas for the gravitational force. In section \ref{weak}, we introduce the background Lorentz transformation technique that we rely on in sections \ref{four} and \ref{five} to find expressions for forces in the two reference frames and compare them to resolve the paradox.

\section{Implied gravitational force in any reference frame} \label{coo}
The 4-vector acceleration $A^\alpha$ is the effect of a real (physical) force, generated for example by electromagnetism.
Its components are related to the 4-velocity elements $U^\alpha$ via a covariant derivative with respect to proper time $\tau$:
\begin{equation}
\label{A}
A^\alpha = \frac{DU^\alpha}{d\tau} = \frac{dU^\alpha}{d\tau} + \Gamma^\alpha_{\mu\nu} U^\mu U^\nu
\end{equation}

where Greek indices run over spacetime coordinates $\{0,1,2,3\}$, and $\Gamma^\alpha_{\mu\nu}$ are the Christoffel symbols that depend on the coordinates specified by our choice of reference frame. 
A force measured in any frame is a 3-vector that depends on the coordinates:
\begin{equation}
F^i = \frac{dP^i}{dt} = \frac{d (m U^i)}{dt} = \frac{m}{\gamma_u} \frac{dU^i}{d\tau}
\end{equation}
where $i \in \{1,2,3\}$ runs over spatial coordinates, $m$ is the rest mass, and we change $dt$ to the body's proper time via: $dt = \gamma_u d\tau$. $\gamma_u$ is a function of $u$ and metric components, with $u$ being the 3-velocity of the body in the chosen frame. In Minkowski spacetime, $\gamma_u =  1/\sqrt{1 - u^2}$.
In a non-inertial frame, gravity appears as an inertial (fictitious) force due to the term containing the Christoffel symbols. If we consider a body free from real forces, then the 4-acceleration components vanish, and we find using eq. (\ref{A}) the force observed in that frame to be:
\begin{equation}
\label{For} F^i = \frac{m}{\gamma_u}
(-\Gamma^i_{\mu\nu} U^\mu U^\nu)
\end{equation}
With the aid of eq. (\ref{For}), all we need to do to find the gravitational force in any frame is to find the metric components in that frame. This frees us from any force transformation assumptions.
\section{Weak field approximation and the background Lorentz transformation} \label{weak}
Earth's gravity near the surface can be modeled using a 'nearly' flat spacetime \cite{Sch}. This can be expressed in the rest frame of Earth, $R$, and in a small volume near Earth's surface as:
\begin{equation}
    g_{\alpha \beta} = \eta_{\alpha \beta} + h_{\alpha \beta} 
\end{equation}
where $\eta_{\alpha \beta} = \diag \{-1,1,1,1\}$ are Minkowski metric components, and $ | h_{\alpha \beta} | \ll 1$.
Using the condition that general relativity must, in Earth's weak gravitational field, reproduce the same predictions of Newtonian gravity, one finds \cite{Sch} that all off-diagonal components of $h_{\alpha \beta}$ vanish, and:
\begin{equation}
    h_{00} = h_{11} = h_{22} = h_{33} = -2\phi
\end{equation}
where $\phi$ is the Newtonian gravitational potential. We need only consider a small region close to Earth's surface, and therefore we can use the approximation $\phi = g z \ll 1$, where $g$ is a constant gravitational acceleration, $z$ is the vertical coordinate, and we set the speed of light $c=1$.

Then, we need to find what the metric components will look like in the coordinates of a reference frame $\bar{R}$ in horizontal motion at velocity $\boldsymbol{v} = v \boldsymbol{\hat{x}}$ with respect to $R$. Note that we will use this general frame $\bar{R}$ to deduce formulas in sections \ref{four} and \ref{five} for both the Earth's rest frame and the Submarine's frame.
All we need in the small region of space we consider is a local coordinate transformation. In the case of nearly flat spacetime, it turns out to be the familiar Lorentz transformation \cite{Sch}.
Briefly, we know that Lorentz contraction in the $\boldsymbol{\hat{x}}$ direction must result in the coordinate transformation: $\bar{x} = \gamma(x - v t)$, where $\gamma =  1/\sqrt{1 - v^2}$. In turn, the $x$ coordinate can be expressed in terms of $\bar{x}$ and $\bar{t}$ by changing the sign of $v$: $x = \gamma(\bar{x} + v \bar{t})$.
Combining these two equations can give us $\bar{t}$ in terms of $x$ and $t$. As for the spatial coordinates perpendicular to the motion, they must be the same in both coordinate systems: $\bar{y} = y$, $\bar{z} = z$.
We can perform a background Lorentz transformation for a boost of velocity $v$ in the $x$ direction with $x^{\bar{\alpha}} = \Lambda^{\bar{\alpha}}_{\;\;\beta} x^\beta$ and the usual Lorentz transformation matrix
\begin{equation}
    \Lambda^{\bar{\alpha}}_{\;\;\beta} =
    \begin{bmatrix}
    \gamma&-v\gamma&0&0\\
    -v\gamma&\gamma&0&0\\
    0&0&1&0\\
    0&0&0&1
    \end{bmatrix}
\end{equation}
takes the background spacetime coordinates to the reference frame $\bar{R}$ moving at velocity $\boldsymbol{v} = v\boldsymbol{\hat{x}}$ with respect to $R$. We consider Earth's surface curvature to be negligible in the small region we study, so $x$ is a horizontal Cartesian coordinate. We can see that metric components transform like
\begin{gather}
    g_{\bar{\alpha}\bar{\beta}} = \Lambda^{\mu}_{\;\;\bar{\alpha}} \Lambda^{\nu}_{\;\;\bar{\beta}} g_{\mu\nu} = \Lambda^{\mu}_{\;\;\bar{\alpha}} \Lambda^{\nu}_{\;\;\bar{\beta}} \eta_{\mu\nu} + \Lambda^{\mu}_{\;\;\bar{\alpha}} \Lambda^{\nu}_{\;\;\bar{\beta}} h_{\mu\nu}
\end{gather}
The matrix $\Lambda^{\mu}_{\;\;\bar{\alpha}}$ is exactly like the matrix $\Lambda^{\bar{\alpha}}_{\;\;\beta}$ except with $v$ changed to $-v$ \cite{Sch}.
Now $g_{\bar{\alpha}\bar{\beta}}$ expresses how the metric would look like from the point of view of an observer moving at velocity $\boldsymbol{v} = v\boldsymbol{\hat{x}}$ with respect to $R$.
Knowing that the Lorentz transformation has the property $\eta_{\bar{\alpha}\bar{\beta}} = \Lambda^{\mu}_{\;\;\bar{\alpha}} \Lambda^{\nu}_{\;\;\bar{\beta}} \eta_{\mu\nu}$, we can write
\begin{gather}
    g_{\bar{\alpha}\bar{\beta}} = \eta_{\bar{\alpha}\bar{\beta}} + h_{\bar{\alpha}\bar{\beta}}\\
    h_{\bar{\alpha}\bar{\beta}} \equiv  \Lambda^{\mu}_{\;\;\bar{\alpha}} \Lambda^{\nu}_{\;\;\bar{\beta}} h_{\mu\nu}
\end{gather}
where the latter equation is a definition. We see that $h_{\mu\nu}$ under a background Lorentz transformation behaves like a tensor in special relativity. In effect, our setup describes a theory of a symmetric tensor field  $h_{\mu\nu}$ propagating on a flat background spacetime. We find for its components in $\bar{R}$:
\begin{equation}
    h_{\bar{\alpha}\bar{\beta}}= -2gz
    \begin{bmatrix}
    \frac{1+v^2}{1-v^2}&\frac{2v}{1-v^2}&0&0\\
    \frac{2v}{1-v^2}&\frac{1+v^2}{1-v^2}&0&0\\
    0&0&1&0\\
    0&0&0&1
    \end{bmatrix}
\end{equation}

Even though $\gamma$ can be large, we have the freedom to assign $z \approx 0$ without loss of generality and maintain $|h_{\bar{\alpha}\bar{\beta}}| \ll 1$.
To first order in $h_{\mu\nu}$, we can approximate $g^{\mu\nu} \approx \eta^{\mu\nu} - h^{\mu\nu}$ and write:
\begin{gather}
    \Gamma^\lambda_{\mu\nu} = \frac{1}{2} g^{\lambda \sigma} (\partial_\mu g_{\sigma \nu} + \partial_\nu g_{\sigma \mu} - \partial_\sigma g_{\mu \nu})\\
    \approx \frac{1}{2} \eta^{\lambda \sigma} (\partial_\mu h_{\sigma \nu} + \partial_\nu h_{\sigma \mu} - \partial_\sigma h_{\mu \nu})
\end{gather}
We find the only non-zero Christoffel symbols to be:
\begin{subequations}
\begin{gather}
\Gamma^{\bar{0}}_{\bar{0}\bar{3}} = \Gamma^{\bar{0}}_{\bar{3}\bar{0}} = \Gamma^{\bar{3}}_{\bar{0}\bar{0}} = \Gamma^{\bar{3}}_{\bar{1}\bar{1}} = -\Gamma^{\bar{1}}_{\bar{1}\bar{3}} = -\Gamma^{\bar{1}}_{\bar{3}\bar{1}} = g\gamma^2(1+v^2)\\
\Gamma^{\bar{0}}_{\bar{1}\bar{3}} = \Gamma^{\bar{0}}_{\bar{3}\bar{1}} = \Gamma^{\bar{3}}_{\bar{0}\bar{1}} = \Gamma^{\bar{3}}_{\bar{1}\bar{0}}
= - \Gamma^{\bar{1}}_{\bar{0}\bar{3}} = - \Gamma^{\bar{1}}_{\bar{3}\bar{0}} = 2g\gamma^2 v\\
\Gamma^{\bar{3}}_{\bar{2}\bar{2}} = -\Gamma^{\bar{3}}_{\bar{3}\bar{3}} = -\Gamma^{\bar{2}}_{\bar{3}\bar{2}} = -\Gamma^{\bar{2}}_{\bar{2}\bar{3}} = g
\end{gather}
\end{subequations}
We can now
input these Christoffel symbols into (\ref{For}) to find a general expression for the
gravitational force acting on a body with 4-velocity $\bar{U}^\alpha = (\gamma_{\bar{u}},
\gamma_{\bar{u}} \bar{u}_x, \gamma_{\bar{u}} \bar{u}_y, \gamma_{\bar{u}} \bar{u}_z)$ in a reference frame $\bar{R}$ which is in a small region near Earth's surface and moves at velocity $\boldsymbol{v} = v\boldsymbol{\hat{x}}$ with respect to Earth.
Note that $\gamma_u = 1/\frac{d\tau}{dt} = 1/ \sqrt{1+2gz -(1-2gz)u^2} \approx \sqrt{1 -u^2} \approx \gamma_{\bar{u}}$ within our approximation (or by setting $z=0$).

Since we wish to study the implied gravitational force on the submarine and on water, and since they both have zero velocity components in the $\boldsymbol{\hat{y}}$ and $\boldsymbol{\hat{z}}$ directions, the only relevant Christoffel symbols are found to be $\Gamma^{\bar{3}}_{\bar{0}\bar{0}}$, $\Gamma^{\bar{3}}_{\bar{1}\bar{0}}$, $\Gamma^{\bar{3}}_{\bar{0}\bar{1}}$, and $\Gamma^{\bar{3}}_{\bar{1}\bar{1}}$. This means that the force is in the $\boldsymbol{\hat{z}}$ direction, and is given by:
\begin{equation}
\label{F}
\boldsymbol{\bar{F}} = -\gamma_{\bar{u}} \gamma^2 ((1+v^2)(1+\bar{u}_x^2)+4v\bar{u}_x) mg \boldsymbol{\hat{z}}
\end{equation}

\begin{figure*}
    \centering
    \includegraphics[width=17cm]{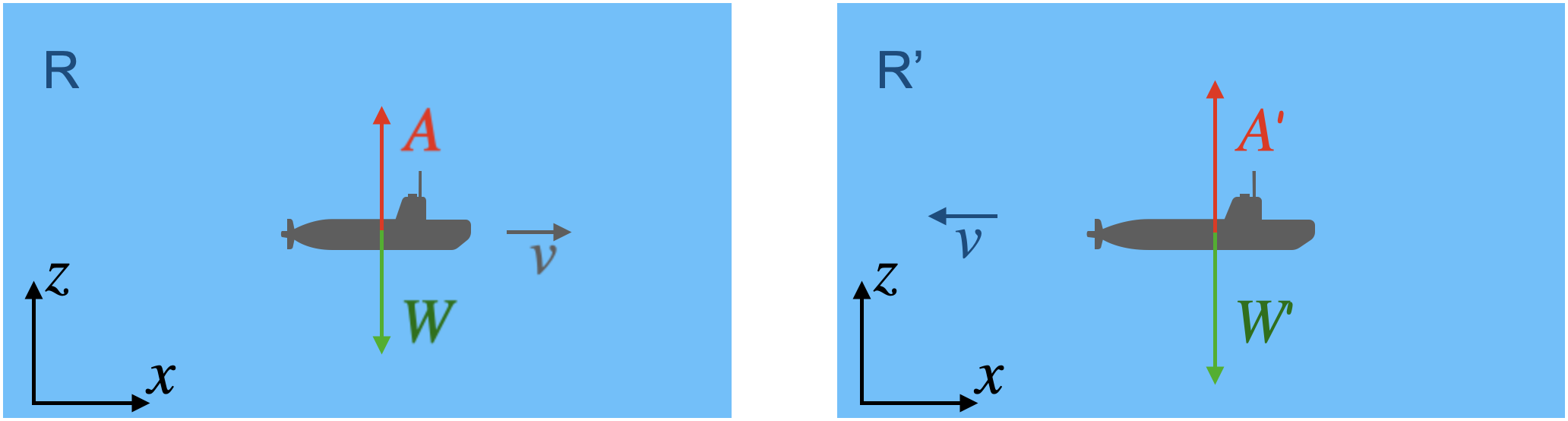}
    \caption{\textit{Submarine paradox sketch}: In $R$, A submarine moves with velocity $\boldsymbol{u_s} = v \boldsymbol{\hat{x}}$ while the water is at rest. The density of the water is adjusted to the moving submarine's density such that the Archimedes force $A$ cancels the gravitational force $W$ and the submarine is in equilibrium.
    In $R'$,
    the submarine is at rest while the water moves with velocity $\boldsymbol{u'_w} = -v \boldsymbol{\hat{x}}$. Due to Lorentz length effects, the submarine's density becomes lower than the water’s. Nevertheless, a proper transformation of the background metric shows that the Archimedes force $A'$ still cancels the gravitational force $W'$ and the submarine remains in equilibrium.}
    \label{fig:my_label}
\end{figure*}

\section{Forces in R} \label{four}
we start in Earth's rest frame where $\boldsymbol{v} = 0$ and eq. (\ref{F}) simplifies to:
\begin{equation}
\label{FR}
\boldsymbol{F} = -\gamma_u (1+u_x^2) mg \boldsymbol{\hat{z}}
\end{equation}
Notice that expression (\ref{FR}) differs from the force obtained by Vieira \cite{Vie} and Supplee \cite{Sup}, as both rely on a metric valid ``in the limiting case of small velocities \cite{Landau}" and formulas derived from it. That metric only differs from Minkowski's in $g_{00} = -1 -2\phi$. In fact, if we redo our calculations starting with that metric, we get $\Gamma^{\bar{3}}_{\bar{1}\bar{1}} = 0$, leading to the expressions they write for the force. Veiera was able to resolve the paradox because, in principle, the paradox must still be resolvable under the assumption of small velocity and in any metric. However, our goal is a solution for relativistic velocities, and the small velocity limit is incorrect.

Having obtained the expression for the force, we can now use it to find the weight of the submarine $\boldsymbol{W_s}$ and the Archimedes force $\boldsymbol{A}$. For this, we note that in $R$, the water is at rest while the submarine is moving at $\boldsymbol{u_s} = v\boldsymbol{\hat{x}}$. We find:
\begin{equation}
\boldsymbol{W_s} = -\gamma (1+v^2) m_s g \boldsymbol{\hat{z}} = -\gamma (1+v^2) \rho_s V_s g \boldsymbol{\hat{z}}
\end{equation}
where $m_s$ is the rest mass of the submarine, $\rho_s$ is its density in $R$, and $V_s$ is its volume in $R$.
As for the Archimedes force, we first find the gravitational force $\boldsymbol{F_w}$ on a water element of rest mass $m_w$, rest volume $V_w$, and rest density $\rho_w$ from (\ref{FR}) with $u=0$. Dividing it by $V_w$ we then compute the force's volume density, which is equal to the pressure gradient, and perform a volume integral over the submarine's volume in $R$ to find the Archimedes force:
\begin{gather}
\boldsymbol{F_w} = -m_w g \boldsymbol{\hat{z}}\\
\boldsymbol{\nabla} p = \boldsymbol{f_w} = \boldsymbol{F_w}/V_w = - \rho_w g \boldsymbol{\hat{z}}\\
\boldsymbol{A} =  \int_{V_s} (\boldsymbol{-\nabla} p) dV = \rho_w V_s g \boldsymbol{\hat{z}}
\end{gather}

As illustrated in fig. (\ref{fig:my_label}), in order to resolve the paradox we can find the equilibrium condition in $R$ and show that it leads to equilibrium in the submarine's reference frame $R'$.
We impose equilibrium by setting $\boldsymbol{A} = -\boldsymbol{W_s}$. For this, we require:
\begin{equation}
\label{rho}
\rho_w = \gamma(1+v^2) \rho_s
\end{equation}
\section{Forces in R$'$} \label{five}
The submarine's frame moves at velocity $\boldsymbol{v} = v\boldsymbol{\hat{x}}$ with respect to Earth, which means that the metric undergoes the background Lorentz transformation we performed to get equation  (\ref{F}). We find:
\begin{equation}
\label{FR'}
\boldsymbol{F'} =  -\gamma_{u'} \gamma^2 ((1+v^2)(1+u'^{2}_x) +4v u'_x)mg\boldsymbol{\hat{z}}
\end{equation}
We can now compute the weight of the submarine $\boldsymbol{W'_s}$ and the Archimedes force $\boldsymbol{A'}$ in $R'$. The submarine here is at rest, so we substitute $\boldsymbol{u'} =0$ to find:
\begin{equation}
\boldsymbol{W'_s} = -\gamma^2 (1+v^2) m_s g \boldsymbol{\hat{z}}
= -\gamma^2 (1+v^2) \rho_s V_s g \boldsymbol{\hat{z}}
\end{equation}
For the Archimedes force, we follow the same steps as before, but now the water is moving at velocity $\boldsymbol{u} = -v\boldsymbol{\hat{x}}$, and we use the densities and volumes in $R'$.
\begin{gather}
\boldsymbol{F'_w} = -\gamma^3 m_w g ((1+v^2)^2 -4v^2) \boldsymbol{\hat{z}}
= -\frac{1}{\gamma} m_w g \boldsymbol{\hat{z}}\\
\boldsymbol{\nabla'} p' = \boldsymbol{f'_w} = \boldsymbol{F'_w}/V'_w = - \frac{1}{\gamma} \rho'_w g \boldsymbol{\hat{z}}\\
\boldsymbol{A'} =  \int_{V'_s} (\boldsymbol{-\nabla'} p') dV = \frac{1}{\gamma} \rho'_w V'_s g \boldsymbol{\hat{z}}
\end{gather}

Knowing that due to Lorentz contraction we have $\rho'_w = \gamma \rho_w$ and $V'_s = \gamma V_s$, and recalling the equilibrium condition in $R$ from eq. (\ref{rho}) we find:
\begin{equation}
\boldsymbol{A'} = \gamma \rho_w V_s g \boldsymbol{\hat{z}}
= \gamma^2(1+v^2) \rho_s V_s g \boldsymbol{\hat{z}} = -\boldsymbol{W'_s}
\end{equation}

Thus, we have equilibrium in $R'$ as well and the submarine paradox is resolved.

\section{Conclusion}
Reviewing our results, we see that in the reference frame of the submarine, both the weight of the submarine and the buoyancy force increase by a factor of $\gamma$. Therefore, if the submarine sinks in one frame, it must sink in the other.
Of course, this is not the only way to resolve the paradox in general relativity \cite{Mat}, \cite{Ric}. However, with this solution we find a new and fundamental perspective: the paradox arises when we mistakenly consider the metric components of an observer moving with respect to earth to be identical to that of observers at rest. A resolution to the paradox then lies in the application of the background Lorentz transformation technique to relate metric components of two observers moving at a constant velocity with respect to one another.
Another interpretation of the result within the nearly flat spacetime approximation is also possible: one splits the metric into a Minkowski metric and a non-Minkowskian part which can be interpreted as a tensor field. This field generates a Newtonian force field, and its Lorentz transformation properties can quantitatively account for the consistency of physics across reference frames.

\acknowledgments
I am grateful to Ewan Stewart for supervising an initial version of this work for my BSc thesis at KAIST.

\end{document}